\documentclass[12pt]{article}
\usepackage{amsmath}
\usepackage{amsfonts}   
\usepackage{amssymb}    

\usepackage{epigraph}
\usepackage{color}
\usepackage{epsfig}
\usepackage{hyperref}
\usepackage{cite}
\usepackage[english]{babel}
\usepackage{graphicx}
\usepackage{epstopdf} 
\usepackage{url}
\usepackage{breakurl}

\def\la{\lambda}

\def\si{\sigma}

\def\beq{\begin{equation}}
\def\eeq{\end{equation}}

\def\la{\langle}
\def\ra{\rangle}

\begin{document}
\title{Alternate bearing and possibile long-range communication of {\it Olea europaea}\\} 
\author{\textsc{Sergei Esipov}\\
\it{Quant Isle Ltd., New York, USA}\\
\it{and}\\
\textsc{Clara Salue\~na}\\
\it{Department of Mechanical Engineering,}\\
\it{Av. Pa\"isos Catalans 26, 43007 Tarragona, Spain}}
\date{}
\maketitle
\vfill
\begin{center}
{\large July 28, 2018}
\end{center}
\begin{abstract}



{Spatio-temporal analysis typically performed in horticulture and in statistical physics reveals persistent correlations of olive yields which range depend on the {\it size} of the averaging regions. Mapping spatially correlated regions unveils areas which resemble historical spread of {\it Olea europaea}. These yield patterns are remarkable given the intensive nature of modern agriculture, and cannot be attributed to weather or pollen indices due to inability of these variables to properly predict yields, and their different correlation patterns. Long-range correlations between yields of olive trees may indicate long-range communications among trees. }

\end{abstract}
\vfill
\newpage


\section{Introduction}

\epigraphrule 0pt
\epigraphwidth 330 pt
\epigraph{Another problem of evaluation, useful when discussing alternation,
seems to have not been quantitatively approached: synchronization of
different plants within a single orchard or of different orchards within a
single region. Such an attempt would provide a basis to evaluate to what
extent external factors (common to a grove, an area, etc.) are dominant 
as against internal factors of trees or factors common to a restricted
area, such as microclimate, soil-rootstock-cultivar relationships, etc.}
{\cite{shoots}, p. 131}

About four decades have passed since the above request for a quantitative model 
of alternate bearing was written, these decades saw an intensifying research 
into the origin and control of alternate bearing  - a phenomenon where crops 
alternate, year after year, in an approximate 'year on - year off' but otherwise 
seemingly random fashion. Olive trees, with their praised fruits, and relatively 
detailed historical records, are similar to many other plants in this regard. A 
year of abundance is usually followed by a year of low crops, and vice versa, 
but not necessarily, and not everywhere. Alterations extend over at least 5-6 
orders of magnitude in space, ranging from a single olive tree (leaving aside 
the branch-to-branch alterations) to scales of hundreds if not thousands of 
kilometers, at times - across a body of water. This phenomenon is known to 
resist human control which routinely includes such drastic measures as 
cultivation of young trees, \footnote{By age anywhere from one to three dozen years, depending on local agricultural 
practice, the trees are 
considered to lose their 'vigor', and are replaced by young trees.} aggressive 
annual pruning, planting trees in a optimized dense grid 
\cite{tous1997planting}, and a wide spectrum of biochemical applications. 

On one side, truly biannual alterations could only be in one of two phases (on 
or off for a given tree in a given year), which would make quickly decaying 
random spatial patterns after spatial averaging were it not for synchronized 
regions which may extend over hundreds of kilometers, and include geographically 
separated trees. On the other side, there is no known clear predictor of yield 
at large scales (except, obviously, the yield itself), while there is no 
shortage of factors considered,  \cite{lavee2007biennial}. Among external 
factors, attempts to use pollen \cite{recio1996olea}, \cite{mazzeo2014amount}, 
and weather-related variables for crop predictions are repeatedly revisited 
\cite{hartmann}, \cite{rallo1993dormancy}, \cite{fornaciari2005yield}, 
\cite{galan2008modeling}, \cite{fatima2014}.  

Below, starting with data analysis, we briefly present a comparison of different 
correlation measures which either have been of could be of interest regarding 
alternate bearing, and argue that these measures cannot be explained 
endogenously, suggesting a potential long-range communication between olive 
trees.  

\section{Spatio-temporal correlations in olive plantations}

In this section we will address the correlations of the surface density of the 
annual olive crops. This variable is termed 'yield' in agriculture. The time 
scale $t$, for yield, $Y_t(x)$ is discrete (integer indices $t$, $t+1$, $t+2$, 
... refer to years), while $x$ are spatial scales. Historical data are seldom 
available at the sub-tree level, but tree-level data are sometimes available at 
research institutions, and public data can be found at county, province and 
country level and can be obtained from government or non-for-profit sources and 
from The Statistics Division of The United Nations \cite{faoun}.  

We first perform a study of the type usually done in statistical physics 
\cite{Landau6}.  Since plants are inherently Malthusian systems, we consider 
correlations of normalized logarithmic yields, \footnote{This allows one to 
focus more on the spatio-temporal relationships rather than on the details of 
the yield magnitude.} 
\beq
\label{eq:ylog}
y_t(x) = \frac{1}{\si(x)}\left[\log Y_t(x) - \la \log Y(x)\ra\right],
\eeq
where $\la ...\ra$ is an average over multiple years (hence, no time index), and 
$\si(x)$ is the standard deviation of the logarithm (again, no time index). The 
average  
\beq
\label{eq:corr1}
\rho_1(t' - t) = \la y_t(x) y_{t'}(x)\ra,
\eeq
is then the correlation coefficient over the lag of $t' - t$ years, called the 
autocorrelation below. Here $t'-t = 1,2, ...$. 
\begin{table}
\begin{center}
    \begin{tabular}{ | l | l | l | l | l | l | l | l |}
    \hline
Location	& Years &	$\rho_1(1)$ & $L$, km	&	$B$	&	$I_B$	&	$S$	&	$I_S$		\\
\hline
Montsi\`{a} 	& 2008-2013 &	$-0.66\pm0.23$	& n/a &	1	&	0.28	&	n/a	&	n/a  		\\
M\'{a}laga	 & 1999-2013 &	$-0.55 \pm 0.18$	 & n/a &	0.77	&	0.17	&	n/a	&	 n/a 	\\
Tarragona & 2006-2013 &	$-0.02 \pm 0.35$ 	& 67 &	0.83	&	0.15	&	0.5 to 1	&	 0.37 to 0.45	\\
Lleida & 2006-2013 &	$-0.08 \pm 0.35$ 	& 248 &	0.5	&	0.18	&	0.5 to 1	&	 0.31 to 0.36	\\
Andalucia	& 1999-2013 &	$-0.31 \pm 0.23$ 	& 189 &	0.61	&	0.18	&	0.5  to 1	&	0.20 to 0.32  		\\
Spain  & 1999-2013	&	$-0.36 \pm 0.16$	 & n/a &	0.5	&	0.11	&	n/a	&	n/a 		\\
World & 1999-2013 	&	$-0.28 \pm 0.24$	 & 603 &	0.52	&	0.03	&	0.52 to 0.55	&	0.40 to 0.57 		\\
\hline
\end{tabular}
\caption{Different measures of alterations. $\rho_1$ is the yield one-year \- 
autocorrelation, \eqref{eq:corr1}, $L$ is the correlation length 
\eqref{eq:corr2}, $B$ - biennuality \eqref{eq:bi1}, $I_B$ - intensity 
\eqref{eq:bi2}, $S$ - synchronization \eqref{eq:S}, $I_S$ - relative amplitude 
of spatial fluctuations \eqref{eq:S2}. At the locations, where we did not have 
subdivision data available, spatial indices are given as  'n/a'. Synchronicity, 
$S$, depends on time, and the minimum and maximum values are given. M\'{a}laga 
is chosen over Jaén because the latter determines the entire Andalusia, and its 
values are close to Andalusian ones.}
\label{tab:bi}
\end{center}
\end{table}
A certain degree of alternation is observed in the time series of yield, and one 
can see in Table \ref{tab:bi} that on average one-year autocorrelations are 
negative, $\rho_1(1) < 0$. The residual noise is not small, and phases of 
biennual patterns change in what seems to be a random fashion. 

\begin{figure}[!h]
  \begin{center}
	\advance\leftskip -0cm
	{\includegraphics
 [width=1.0\textwidth]
{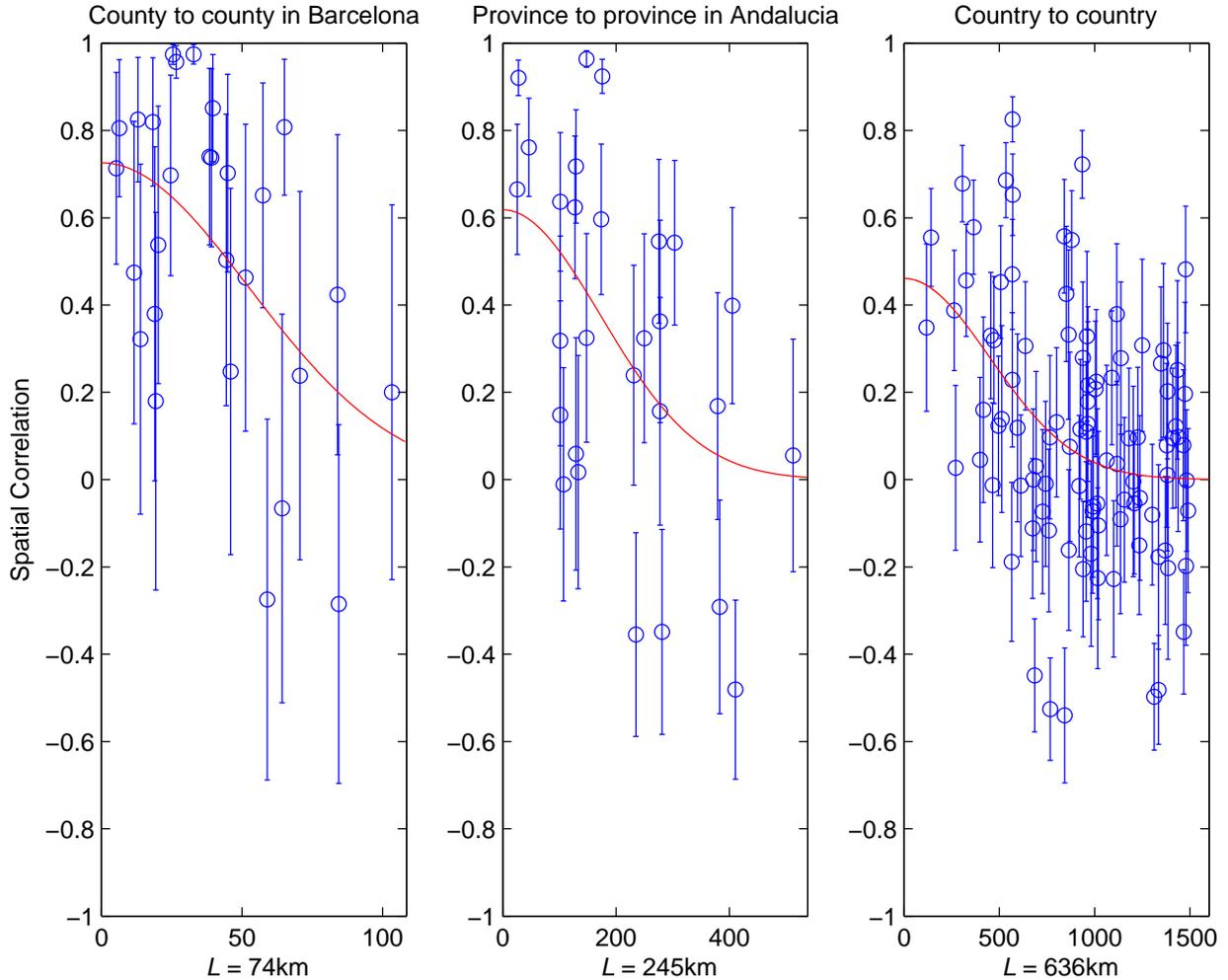}}
  \caption{Decay of spatial correlations $\rho_2$ at different scales. $x$-axis 
is distance in kilometers in all subplots. Vertical lines show error bars. Red 
lines are fits using Eq \eqref{eq:corr2} with correlation length $L$ given 
below each subplots.
}
\label{fig:corr}
\end{center}
\end{figure}

\begin{figure}[!h]
  \begin{center}
	\advance\leftskip -0cm
	{\includegraphics
 [width=1.0\textwidth]
{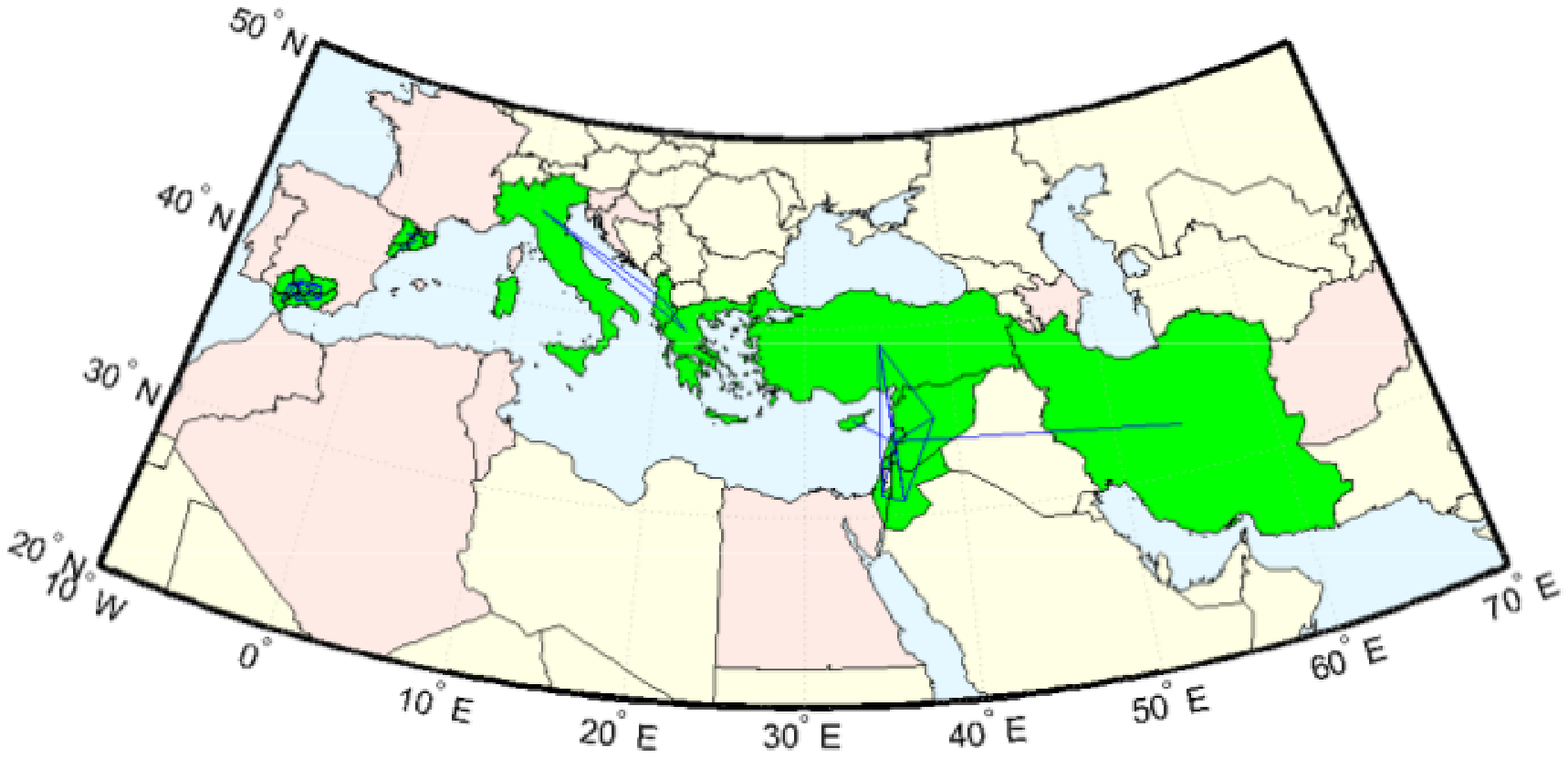}}
  \caption{Green and pink countries submitted official reports to FAOSTAT for at 
least a decade. The countries, which are pairwise correlated with $\rho_2 > 
1/2$, are colored green and connected by straight blue lines. The entire 
countries are colored (even if their olive plantations are far from being 
uniform within country's borders). The correlated countries form two clusters. 
Italy, Albania and Greece form one, and seven countries in the Middle East form 
another cluster. In Spain we show a province-level cluster in Catalonia and 
Andalusia. We do not have data for other provinces in Spain.
}
\label{fig:map}
\end{center}
\end{figure}

The average 
\beq
\label{eq:corr2}
\rho_2(x' - x) = \la y_t(x) y_t(x')\ra
\eeq
is a spatial correlation coefficient over the distance $x' - x$. An important 
property of this correlation is its decay, which we will model by means of a 
single correlation length, 
\beq
\label{eq:corr2a}
\rho_2(x) \propto \exp\left(-x^2/L^2\right).
\eeq
Remarkably, the fitted correlation length, $L$, is found to depend on the 
dataset scale, see Fig\ref{fig:corr}. For the province of Tarragona it is 67 km, 
for Andalusia - 189 km, and for the Europe/Africa/Middle East dataset it is 603 
km, see also Table \ref{tab:bi}. \footnote{Similar phenomena exist in 
turbulence, where fluid velocity correlations depend on the sampling scale, 
because progressively bigger eddies begin to contribute \cite{Landau6}.} The 
largest of these correlations lengths is comparable to the latitudinal span of 
The Mediterranean Sea. 
Scale-dependent spatial correlations of olive yields indicate a presence of 
hierarchical spatial structures. One should then expect that there are subsets 
of locations where correlations decay much slower with distance, as compared to 
what  \eqref{eq:corr2} prescribes, and subsets which decay much faster. And 
indeed, further investigation shows that correlations decay slower with distance 
in Lleida province of Catalonia than in Tarragona, and that C\'{o}rdoba, 
Granada, Ja\'{e}n, and M\'{a}laga act as a single entity in Andalusia, and the 
same can be said, for example, about olive yields in Turkey, Syria, Lebanon, 
Cyprus, et cetera. It is then of interest to obtain a correlation map, which 
gives a visual representation of these connections, see Fig \ref{fig:map}.

\section{Exogenous yield predictors}
A large number of factors which are exogenous to olive trees have been 
considered as yield predictors. These factors include weather (namely, 
temperature, humidity and precipitation) and solar irradiation. As these factors 
are themselves time-dependent, various filtered sums have been suggested as 
yield predictors \cite{hartmann}, \cite{rallo1993dormancy}, 
\cite{fornaciari2005yield}, \cite{galan2008modeling}, \cite{fatima2014}. In 
addition the significance of olive pollen has been recognized, 
\cite{recio1996olea}, \cite{mazzeo2014amount}.  A belief that there might exist 
a very detailed weather-related predictor, along the lines of climate 
definitions and cluster analysis is persistent due to complexity of weather 
records. 
The following paragraph is taken from \cite{fatima2014}, ``The meteorological 
variables considered in the present study were: monthly average of maximum and 
minimum temperatures ($^\circ$C) and relative humidity (\%), monthly accumulated 
precipitation (mm) and evapotranspiration (mm day–1), summer period (21st of 
June to 21st of September) accumulated mean temperature ($^\circ$C) and 
accumulated precipitation (mm), and accumulated precipitation since the 
pre-flowering start date (end of March) until the peak pollination day (mm).''

Indeed, the scales of the largest clusters in Fig\ref{fig:map} are comparable to 
correlation lengths for monthly averages of weather variables used in climate 
definitions. 
Nevertheless, the maps of spatial correlations of pollen, which reflect wind 
patterns, and other weather-based variables look quite different from 
Fig \ref{fig:map} (see below). \footnote{See monthly maps of wind, temperature, 
precipitation, etc available from the International Research Institute for 
Climate and Society, Earth Institute, University of Columbia, 
https://iri.columbia.edu.} 

Since the yield time series which are regressed on these external factors are 
relatively short (they usually cover a couple of decades), it is always possible 
to find a meteorological candidate for yield regression, especially using 
filtered variables (such as a variable exceeding an optimized threshold). While 
the original findings of Hartmann and Porlingis \cite{hartmann} have been 
verified on numerous occasions, it is important to have sufficient out-of-sample 
analysis to avoid statistical overfitting.  

The overall situation with using weather for explaining olive yields could be 
summarized as follows. While the weather and pollen indices continue to evolve 
to 'explain' most recent data, the spatial correlations of the proposed 
exogenous indices do not reflect the same for the yields. At the same time, the 
overall existence of exogenous explanation is not questioned.  

The yield correlation maps, however, resemble those of the olive tree varieties. 
Namely, correlation maps at the lowest level have resemblance to local olive 
varieties, while Fig\ref{fig:map} which shows the largest scale considered here 
resembles the three historical 'inoculation waves' of the olive tree expansion 
in the Mediterranean \cite{besnard2013complex}, based on genetic analysis. 

\section{Bienniality-intensity indices for alternate bearing}

Horticulturists do not consider temporal or spatial correlations introduced 
above. Instead, they rely on frequency (known as $B$, for 'biennuality') and 
severity (known as $I$, for 'intensity') of alterations, defined as 
\cite{shoots} 
\begin{align}
\label{eq:bi1}
B &= \frac{1}{n-2}\sum_{t=2}^{n-1} {\rm sign}\left(Y_{t+1} - Y_t\right) {\rm sign}\left(Y_{t} - Y_{t-1}\right),\\
\label{eq:bi2}
I_B &= \frac{1}{n-1}\sum_{t=2}^n \frac{|Y_t - Y_{t-1}|}{Y_t + Y_{t-1}}
\end{align}
Bienniality indicates how frequently a trend changes, while intensity evaluates 
relative amplitudes of 'swings'. Both quantities supersede correlator 
\eqref{eq:corr1} which origin is in gaussian (i.e. normal) statistics. On the 
other side, bienniality specifically focuses on two-year periods, while many 
olive plantations display more complex patterns (e.g. Ja\'{e}n in 1999-2006 had 
a triennial alternation, rather than biennial).  
Both quantities have their analogs in terms of spatial correlations. Their 
spatial frequency is accessed through the 'synchronization' parameter 
\beq
\label{eq:S}
S = \frac{1}{2} + \left|\frac{1}{2A}\sum_x {\hskip .025 in} {\rm sign}\left[Y_t(x) - Y_{t-1}(x)\right]\right|,
\eeq
which measures relative excess of trend increase or decrease in a given area. 
The value $S = 0.5$ corresponds to complete randomization, while $S = 1$ is full 
synchronization. We have not seen a measure of the amplitude of spatial 
fluctuations (an analog of the intensity above) to be considered in 
horticultural literature, as it is not an exclusive measure of alternative 
bearing. An index of relative spatial fluctuations can be defined as  
\beq
\label{eq:S2}
I_S = \frac{1}{A^2}\sum_{x, x'} {\hskip .025 in} \frac{|Y_t(x) - Y_t(x')|}{Y_t(x) + Y_t(x')}.
\eeq
Examples of numerical evaluation of these parameters can be found in Table 
\ref{tab:bi}. One can see that coarse-graining non only leads to decreasing of 
absolute values of one-year autocorrelation, $\rho_1(1)$, but also (i) the 
biennuality, $B$, is similarly suppressed, while remaining positive, (ii) the 
intensity of alternations, $I_B$, consistently diminishes with a characteristic 
distance of $10^3$ km, (iii) the spatial synchronization, $S$, of fluctuations 
becomes small only at the global level (Lleida or Andalusia, represented by 
counties, could still have $S = 1$), and (iv) the index of relative spatial 
fluctuations, $I_S$, remains large at any level.

Thus, the alternate bearing can be averaged over, but it takes an effort: 
several decades in time and global spatial scales. In other words, for the data 
in Table 1 the alternate bearing can be found on all spatial scales.

\section{Alternate bearing at a grove level}

Data for an experimental grove, where different rootstocks are used, and trees 
are cared for in the same fashion as in agriculture, are particularly valuable, 
since they might contain detailed information about alternate bearing. These  
data require a considerable dedication: one has to maintain experimental groves 
for decades (IRTA). 

The trees in the grove were arranged in a rectangle on a square grid: 10 blocks, 11 trees per block. 

\begin{figure}[!h]
  \begin{center}
	\advance\leftskip -0cm
	{\includegraphics
 [width=1.0\textwidth]
{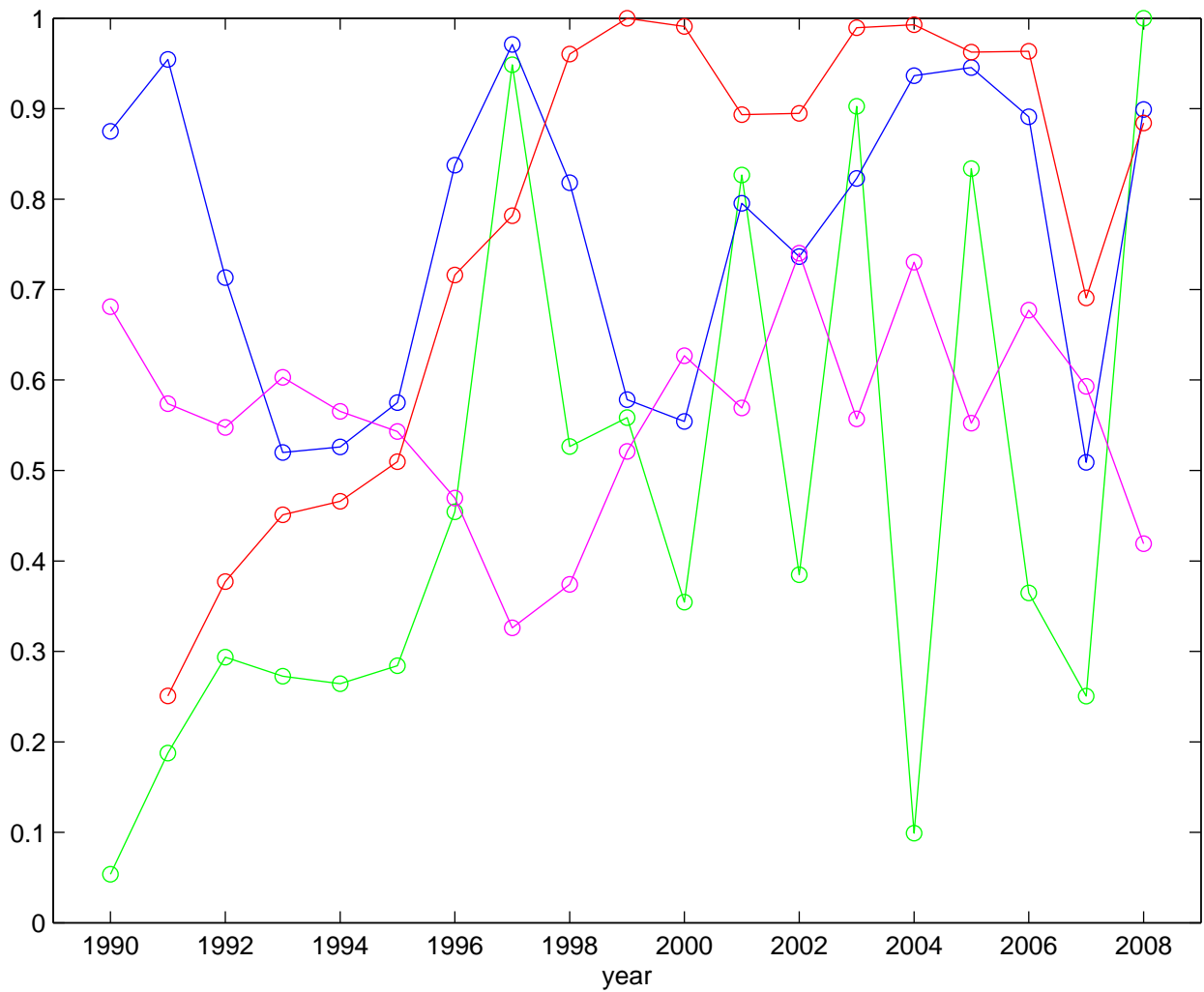}}
  \caption{Time series of normalized net yield (rescaled to maximal yield ever 
observed) - green, synchronization parameter, $S$, - blue and spatial 
synchronization, $I_B$, - magenta, and normalized tree volume - red. } 
\label{fig:ab}
\end{center}
\end{figure}

The overall behavior of the grove yield is shown in Fig \ref{fig:ab}. During the 
first 4 years of bearing (1989-1992, where 1989 is not shown), young trees grew 
exponentially in size and gave increasing yields, while their spatial 
synchronization, $I_S$, reached maximum in 1990, temporal synchronization, $S$ 
reached maximum in 1991, and yield, $Y_t$ reached maximum in 1992, accompanied 
by a decrease in both $S$ and $I_S$. Then, in 1993-1997 we have a consolidation 
phase, in response to increasing pruning, where synchronization in time is fully 
recovered as soon as pruning eased in 1996, while synchronization is space 
plunged to its minimum, allowing trees a semi-individualistic freerun. The 
yields reached a maximum in 1997 and were again pruned (c.f. the red line). In 
response to pruning, alternate bearing was fully developed by year 1999 with 
large negative autocorrelation $\rho_1(1) = -0.8$, and the grove entered an 
alternating regime where synchronization in time remained high, and 
synchronization is space was alternating out-of-phase with the yield (high $Y_t$ 
corresponded to lower $I_S$). In year 2007 trees were severely pruned, and it 
was accompanied by a phase shift of alternate bearing: year 2008 is the first 
even calendar year with a high yield. Overall there is a clear connection 
between alternate bearing and pruning: pruned olive trees enter a consolidation 
stage, regroup and re-synchronize in space to produce high yield in the year 
$t+1$ following excessive pruning in year $t$.  
 
\begin{figure}[!h]
  \begin{center}
	\advance\leftskip -0cm
	{\includegraphics
 [width=1.0\textwidth]
{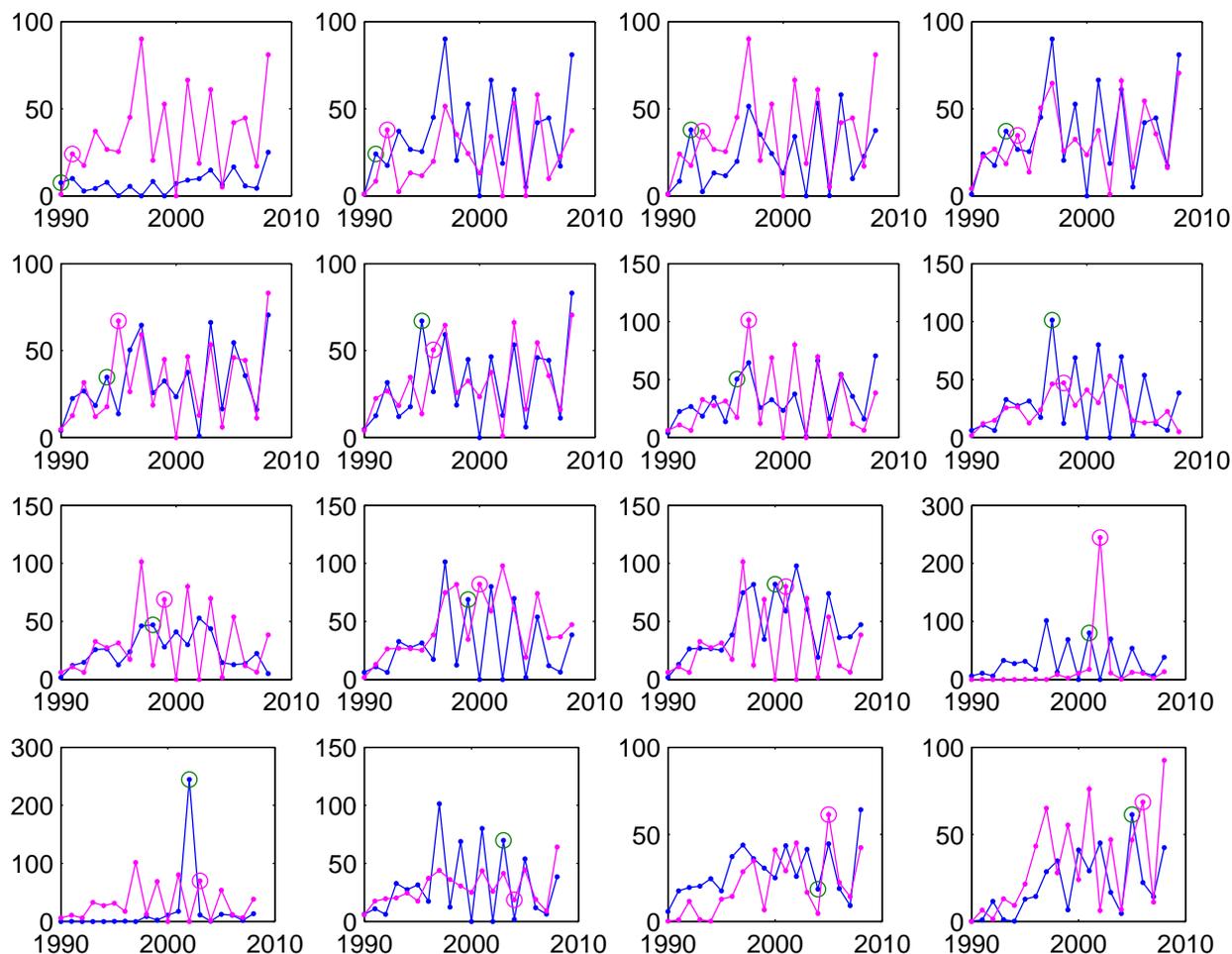}}
  \caption{Synchronization of leaders in two subsequent years. The blue circle 
is the maximal yield in a given year $t$, and blue line connects yields of that 
'blue' tree in all other years. The magenta circle is the maximal yield in the 
following year, $t+1$, and magenta line connects yields of that 'magenta' tree 
in all other years. The $x$-axis is calendar years, and $y$-axis is yield in 
kilograms.} 
\label{fig:leaders}
\end{center}
\end{figure}

To understand whether yield synchronization is a collective phenomenon we 
comparatively time traced the behavior of the individual trees with maximal 
yields. For every two consecutive years, beginning with year 1990, a tree with 
the highest yield was selected, and its yield time series were plotted jointly 
with that of a tree which had the highest yield in the subsequent year 
(beginning with 1991, respectively). In Fig \ref{fig:leaders} we show subsequent 
synchronization of such consecutive leaders. As one can see, in many cases it 
took about two years for ex-leaders to synchronize in terms of alternate 
bearing. Some pairs (i.e. 1998/1999) did not subsequently synchronize, however 
in such infrequent cases the yield of one of the leaders subsequently declined. 
This may indicate that synchronization represents a healthy behavior. Some pairs 
(i.e. 2004/2005) were already synchronized even while being leaders in 
subsequent years. Interestingly, a minority subset of trees can always be 
identified which experiences alternate bearing out-of-phase with the majority. 
The lifetime of this out-of-phase alternate bearing does not exceed a couple of 
years, when the participating minority trees rejoin the quorum, while few new 
majority trees may join the out-of-phase minority. The out-of-phase alternate 
bearing may be used by the tree community for exploration of 'predators' 
appearance during the majority off-years, and whether it is beneficial to switch 
to the opposite on-off phase instead.  

Fig \ref{fig:leaders} shows that synchronization is reestablished in majority of 
cases, and it is not, therefore, reducible to external factors. Indeed, if a 
particular winter chilling were to lead to an increased yield in a given year 
(say, year 1997 when the alternating pattern started), this cannot explain why 
leaders of the consecutive subsequent years had any incentive first to appear 
and then to re-synchronize with the majority. While external factors, such as 
winter chilling, are significant, they cannot substitute for the drivers of 
self-regulation. 

\begin{figure}[!h]
  \begin{center}
	\advance\leftskip -0cm
	{\includegraphics
 [width=1.0\textwidth]
{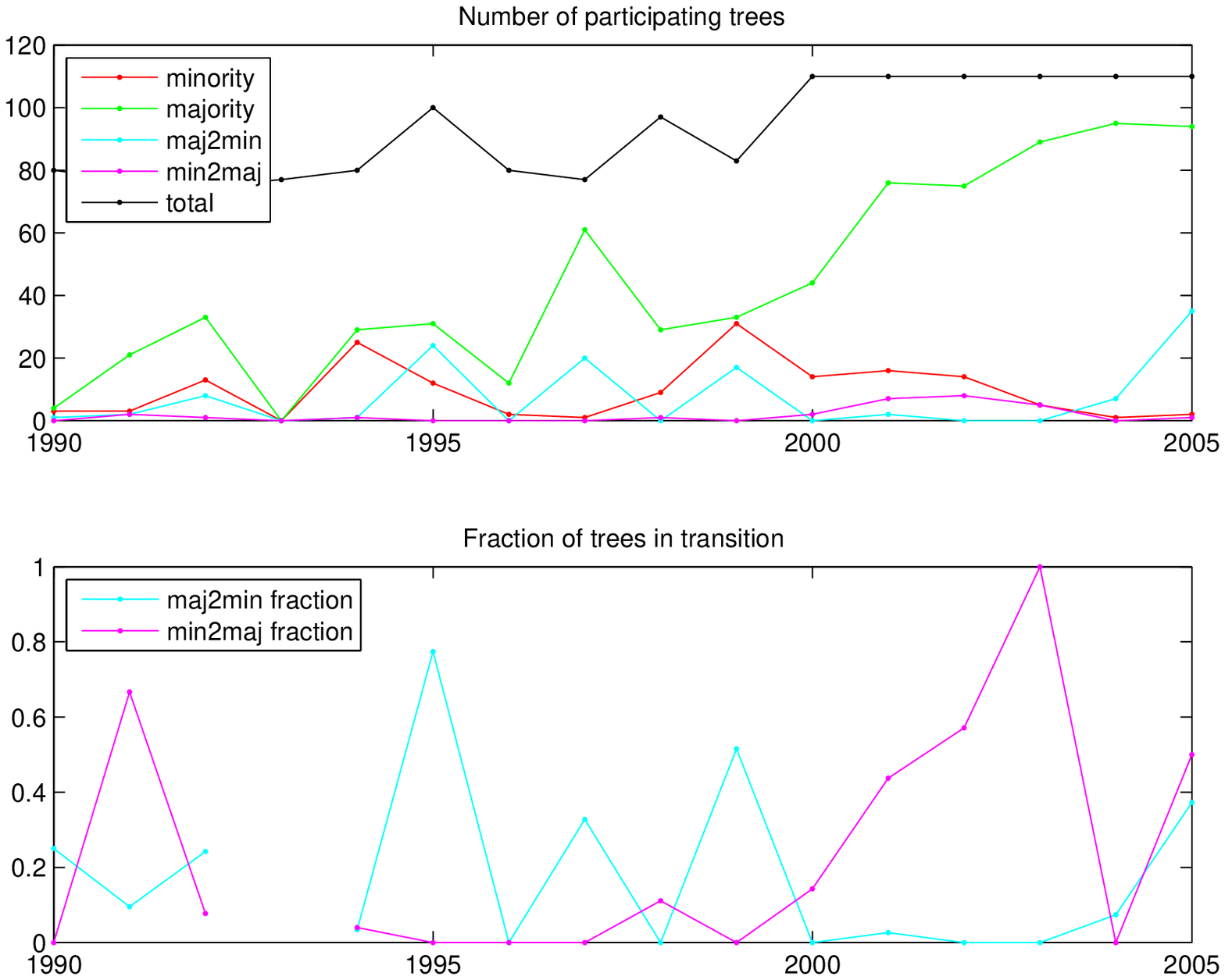}}
  \caption{Top subplot. Number of trees participating in the majority group 
having {\it either} 'on' year ($Y_t > Y_{t-1}$ and $Y_t >  Y_{t+1}$) or 'off' 
year ($Y_t < Y_{t-1}$ and $Y_t <  Y_{t+1}$) is given by the green line, minority 
group - having the opposite year - is given by red line, trees which were 
members of the majority group and ended up as minority 2 years later are shown 
in cyan, and minority group trees synchronized with the majority are shown in 
magenta. Bottom subplot. The fractions obtained by dividing trees in transition 
to the total number of trees in the group they started from. In year $t = 1993$ 
all trees had a yield increase, $Y_{t-1} < Y_t < Y_{t+1}$, so that this year 
cannot be classified as either 'on' or 'off' year for any tree.} 
\label{fig:majmin} 
\end{center}
\end{figure}

Since Fig \ref{fig:leaders} only accounts for leaders, it is indicative of 
synchronization affecting the cases of maximal severity. It is even more 
instructive to see what happens to frequency of participation in the majority 
and minority of trees experiencing alternate bearing. With this in mind in Fig 
\ref{fig:majmin} we plotted the number of trees in both groups, along with the 
number of trees in transition between majority and minority and vice versa.

As one can see, the grove first maintains a comparable number of trees in both 
majority and minority groups (in terms of 'on/off' years), and most trees do not 
participate in either group. Beginning with year 1997, when volume reduction due 
to pruning was (sufficiently?) large, the majority group for the first time 
reaches half of the grove, then the minority is given a chance to increase, but 
none of these adjustments helped to reduce pruning, and the majority is given a 
chance to approach $~85\%$ of the trees. When even this does not help, in 2007, 
following intensive pruning, a switch to minority 'on/off' pattern is performed 
by the entire grove skipping one 'on' year, accumulating resources, and reaching 
the highest net yield ever in 2008. \footnote{This section is a  'local' 
discussion, where communication takes place within the grove. In view of the 
correlation studies in the previous sections we expect that the grove may 
participate in external communications.} 

In terms of spatial synchronization we did not observe any persistent spatial 
patterns. Synchronized trees seemed to occupy random slots, and so did 
out-of-phase trees. In view of high synchronization, this implies that the 
communication between the trees is long-ranged, since, for the randomness to be 
produced, every tree had to have information at least about the average state of 
the entire grove if not about every other tree in the grove. The mechanism of 
this long-ranged communication(s) remains to be identified. Note that this 
mechanism is not reducible to interaction via pollen (which is known to be 
correlated with yield, see above), since fruit bearing requires considerable 
resources to be accumulated in advance. Moreover, in view of the complex 
multistage phenological process unfolding during flowering and fruit bearing, 
any synchronization mechanism has to operate at least on multiple occasions if 
not continuously. 

We conclude that alternate bearing is a collective phenomenon, accompanied by 
syn-chro-ni-za-tion in time, and alternate synchronization in space. Collective 
strategies, followed by the grove, are complex and show a feedback to pruning. 

\section{Conclusion}
Spatial synchronization of alternative bearing is usually attributed to 
exogenous factors, such as pollen or weather indices. We find that the yield 
correlations do not follow the correlations of the 'explanatory' pollen or 
weather variables. The observed synchronization in crop yields, which takes many 
months for trees to bear, is much stronger than what can be explained by pollen 
or weather. Given that trees are adaptive at all stages of their cycle, any 
persistent long-term, long-range synchronization requires similar and very 
frequent if not continuous communication. Given that mature olives still exist 
in the region \cite{lumaret2001plant}, we suggest to consider whether humans 
interact with a plant organism of the size comparable to the Mediterranean 
scales, exceeding the so-called 'Pando' organism of 43.6 ha of aspen trees 
\cite{dewoody2008pando}. Note than inside the Pando organism one also finds more 
than one genetic variate even among aspens. 

\section{Acknowledgment}
The authors are most grateful to colleagues at IRTA (Institute of Agrifood 
Research and Technology, Generalitat de Catalunya, Spain) for access to 
experimental data on olive groves. 

\vfill 
\newpage 
\bibliography{./Olive2}{}
\bibliographystyle{unsrt}
\end{document}